\documentclass[default,iicol]{sn-jnl}

\usepackage{graphicx}%
\usepackage{multirow}%
\usepackage{amsmath,amssymb,amsfonts}%
\usepackage{amsthm}%
\usepackage{mathrsfs}%
\usepackage[title]{appendix}%
\usepackage{xcolor}%
\usepackage{textcomp}%
\usepackage{manyfoot}%
\usepackage{booktabs}%
\usepackage{algorithm}%
\usepackage{algorithmicx}%
\usepackage{algpseudocode}%
\usepackage{listings}%
\usepackage[normalem]{ulem}

\newcommand{\ba}{\begin{eqnarray}}
\newcommand{\ea}{\end{eqnarray}}
\newcommand{\non}{\nonumber}

\theoremstyle{thmstyleone}%
\theoremstyle{thmstyletwo}%

\theoremstyle{thmstylethree}%

\begin{document}

\title[Article Title]{Quantum critical engine at finite temperatures}


\author*[1]{\fnm{Revathy} \sur{B S}}\email{revathy@rrimail.rri.res.in}

\author[2]{\fnm{Victor} \sur{Mukherjee}}\email{mukherjeev@iiserbpr.ac.in}
\author[3]{\fnm{Uma} \sur{Divakaran}}\email{uma@iitpkd.ac.in}


\affil[1]{ \orgname{Raman Research Institute}, \orgaddress{\street{Bengaluru},  \postcode{560080}, \state{Karnataka}, \country{India}}}

\affil[2]{\orgdiv{Department of Physical Sciences}, \orgname{Indian Institute of Science Education and Research Berhampur}, \orgaddress{\street{Berhampur},  \postcode{760010}, \state{Odisha}, \country{India}}}

\affil[3]{\orgdiv{Department of Physics}, \orgname{Indian Institute of Technology Palakkad}, \orgaddress{\street{Palakkad},  \postcode{678623}, \state{Kerala}, \country{India}}}



\abstract{We construct a quantum critical Otto engine that is powered by finite temperature baths. 
We show that the work output of the engine 
shows universal power law behavior that depends on 
the critical 
exponents of the working medium, as well as on the temperature of 
the cold bath. Furthermore, higher temperatures of the cold bath allows the engine to approach the limit of adiabatic operation for smaller values of the time period, while the corresponding power shows a maximum at an intermediate value of the cold bath temperature. These counterintuitive results stems from thermal excitations dominating the dynamics at higher temperatures.}

\keywords{Quantum heat engines, quantum phase transitions, Kibble Zurek scaling}



\maketitle

\section{Introduction}\label{sec1}

Quantum thermodynamics is a rapidly progressing field aimed at 
understanding the thermodynamics at the quantum level 
\cite{gemmer2009quantum, Alicki2018}. The field is getting much 
attention from the scientific community not just because it 
provides a link between quantum mechanics and 
thermodynamics but also because it aids in the development of 
nanoscale devices that aims to harness the potential benefits of 
the ``quantumness" in their working medium (WM) \cite{sai_qt}. 
The recent advances in the experiments such 
as trapped ions or ultracold atoms \cite{rossnagel16a,PhysRevLett.123.080602}, using 
NMR techniques \cite{Peterson18}, nitrogen vacancy centres in diamond   \cite{klatzow19experimental}
have made the realization of quantum devices possible. Among the various quantum devices studied, for example, quantum refrigerators \cite{Abah_2016, hartmann2020multi}, quantum batteries \cite{PhysRevA.97.022106, Campaioli2018QuantumB}, quantum sensors \cite{montenegro23quantum}, quantum thermal transistors \cite{joulain16quantum, PhysRevE.106.024110}, etc., our work focusses on quantum heat engines that follow the quantum Otto cycle \cite{PhysRevE.76.031105, e19040136, PhysRevE.102.012138, Solfanelli_2023, PhysRevB.109.024310, free_fermionicqhe}. 

The effect of phase transitions in quantum engines has been studied 
in Refs. \cite{campisi2016power, Fogarty_2021, 10.1088/1367-2630/ac963b, PhysRevE.96.022143, interaction_chen, PhysRevResearch.2.043247, e24101458, revathy2024improving}. 
While some of these studies concentrated on how to improve the 
performance of quantum engines with respect to its efficiency and 
power \cite{e24101458, revathy2024improving}, some focussed on showcasing
 universality in their working \cite{PhysRevResearch.2.043247, interaction_chen}. 
For instance, in Ref. \cite{PhysRevResearch.2.043247}, the authors 
showed that the work output of the engine up to an additive 
constant follows a universal power law governed by the driving speed 
with which the quantum critical points are crossed, where the exponent of the power 
law is determined by the critical points (CP) crossed. However, they 
considered the case  of relaxing bath that
 takes the WM close to its ground state so that Kibble Zurek mechanism becomes relevant.  
On the other hand, in this paper, we specifically use thermal baths
at different temperatures and study the effect of thermal excitations on the scalings of 
the work output. As discussed before, we 
prepare a many body quantum heat engine using a free fermionic 
model as WM that undergoes a quantum phase transition. 
We describe the details of the free fermionic model in Section \ref{sec2}, 
elaborate on the many body quantum Otto cycle that the engine follows in Section \ref{sec3}, followed by the universal scalings shown by the engine in Section \ref{sec4}. We demonstrate our results using the transverse field Ising model in Section \ref{sec5} and finally conclude in Section \ref{sec6}.

\section{Free fermionic model}\label{sec2}
For a translationally invariant system, free fermionic model can be described by the Hamiltonian
\begin{equation}
H = \sum_k \psi_k^{\dagger} H_k \psi_k,
\end{equation}
with $H_k$ taking the form
\begin{equation}
H_k = (\alpha + m_k) \sigma^z + n_k \sigma^x.
\end{equation}
Here, $\sigma^i$ ($i = x, y, z$) are the Pauli matrices, $\psi_k = ( c_{1k}, c_{2k})$ where $c_{jk} (j = 1,2)$ are the fermionic 
operators corresponding to the $k$-th momentum mode. 
The parameters $\alpha, m_k$ and $n_k$ depend on the specific model that one works on, say Ising \cite{dziarmaga05dynamics}, X-Y \cite{mukherjee07quenching} or Kitaev model \cite{sengupta08exact}. 
The energy gap between the ground state and the first excited state is given by 
$\Delta_k = 2 \sqrt{(\alpha + m_k)^2 + |n_k|^2}$. This Hamiltonian shows a quantum phase 
transition at the quantum critical point where the energy gap $\Delta_k$ vanishes for 
the critical mode $k= k_c$ for certain combinations of $\alpha,~ m_k$ and $n_k$.

With each momentum mode being independent and non-interacting, we can write the density 
matrix of the system as
\begin{equation}
\rho = \otimes_k \rho_k,
\end{equation}
where $\rho_k$ is written in the basis $|0_{1k}, 0_{2k} \rangle, |1_{1k}, 0_{2k} \rangle,
 |0_{1k}, 1_{2k} \rangle$ and $|1_{1k}, 1_{2k} \rangle$ so that the first index corresponds to 
 presence (1) or absence (0) of $c_{1k}$ fermions, which is also the case for second index related to $c_{2k}$ fermions. 
It is to be noted that since the non-unitary dynamics mixes all four basis, we need
to rewrite the Hamiltonian $H_k$ in these four basis leading to \cite{Keck_2017, bandyopadhyay2018exploring}
\begin{equation}
H_{k} = \begin{bmatrix}
 (\alpha +m_k) & 0 & 0 & n_k \\ 0 & 0 & 0 & 0\\ 0 & 0 & 0 & 0\\ n_k & 0 & 0 & -(\alpha + m_k)
\end{bmatrix},
\end{equation} 
whose eigenvalues are $-\epsilon_k, 0, 0, \epsilon_k$ where 
$\epsilon_k = \Delta_k/2=\sqrt{(\alpha + m_k)^2 + |n_k|^2}$.

\section{Many body quantum Otto cycle}\label{sec3}
We now describe the quantum Otto cycle (QOC) which consist of four strokes (also
shown in Fig.\ref{fig_cycle}):

\begin{figure}[h]%
\centering
\includegraphics[scale=.4]{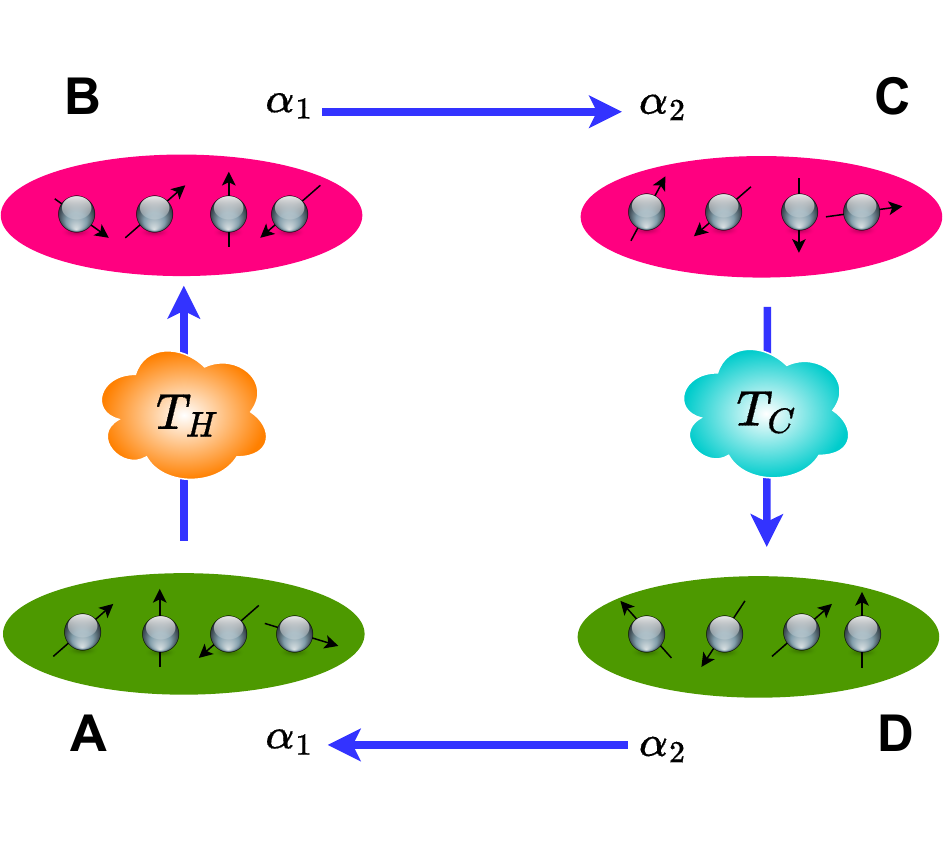}
\caption{Schematic diagram of a quantum Otto cycle}\label{fig_cycle}
\end{figure}

\begin{itemize}
\item[(i)]\underline{Stroke \textbf{A $\rightarrow$ B}}: The WM 
with parameter $\alpha = \alpha_{1}$ is connected to the hot bath 
at a temperature $T_{H}$ for a time $\tau_H$ so that it reaches 
the thermal state at \textbf{B} given by

\begin{equation}\label{eqn_rhoB}
\rho_{k}^B = \begin{bmatrix}
 \frac{e^{\beta_H \epsilon_k}}{Z_H} & 0 & 0 & 0 \\ 0 & \frac{1}{Z_H} & 0 & 0\\ 0 & 0 & \frac{1}{Z_H} & 0\\ 0 & 0 & 0 & \frac{e^{- \beta_H \epsilon_k}}{Z_H}
\end{bmatrix}.
\end{equation} 
Here $\beta_H = \frac{1}{k_{B} T_H}$, (we have set $k_B = 1$ 
throughout this article) and $Z_H = 2 + e^{\beta_H \epsilon_k} + e^{-\beta_H \epsilon_k}$ is the partition function for each mode $k$ with $\epsilon_k$
as the energy when $\alpha=\alpha_1$.
The energy exchanged in this stroke is denoted by $\mathcal{Q}_{in}$.\\

\item[(ii)]\underline{Stroke \textbf{B $\rightarrow$ C}}: The WM is disconnected from the hot bath and $\alpha$ is changed from $\alpha_{1}$ to $\alpha_{2}$ using the driving protocol, 
\begin{equation}
\alpha(t) = \alpha_{1} + (\alpha_{2} - \alpha_{1})(\frac{t}{\tau_1}), \hspace{0.3cm} t \in [0, \tau_1].
\end{equation}
The evolution being a unitary evolution is given by the von-Neumann equation of motion:
\begin{equation}
\frac{d\rho_k}{dt} = -i [H_k, \rho_k].
\end{equation}
In this work, we shall focus on $\alpha_2 = \alpha_c$, the critical value, for the 
reasons that will be explained later.\\

\item[(iii)]\underline{Stroke \textbf{C $\rightarrow$ D}}: The WM 
with  $\alpha=\alpha_{2}$ is next connected to the cold bath  
at a temperature $T_{C}$ till $\tau_C$ so that it reaches the thermal 
state at \textbf{D} given by
\begin{equation}
\rho_{k}^D = \begin{bmatrix}
 \frac{e^{\beta_C \epsilon_k}}{Z_C} & 0 & 0 & 0 \\ 0 & \frac{1}{Z_C} & 0 & 0\\ 0 & 0 & \frac{1}{Z_C} & 0\\ 0 & 0 & 0 & \frac{e^{- \beta_C \epsilon_k}}{Z_C}
\end{bmatrix}
\end{equation} 
where $\beta_C = \frac{1}{k_{B} T_C}$ and $\epsilon_k$ is the energy when
$\alpha=\alpha_2$. The energy exchanged in this stroke is denoted as 
$\mathcal{Q}_{out}$.\\

\item[(iv)]\underline{Stroke \textbf{D $\rightarrow$ A}}: In this last stroke, the WM is disconnected from the cold 
bath, and $\alpha$ is changed back to $\alpha_{1}$ from $\alpha_{2}$ using
\begin{equation}
\alpha(t) = \alpha_{2} + (\alpha_{1} - \alpha_{2})(\frac{t-a}{\tau_2}), \hspace{0.3cm} t \in [a, a + \tau_2]
\end{equation}
to reach \textbf{A} through unitary dynamics and thus the cycle repeats.

\end{itemize} 
Energies at the end of each stroke $i$ is calculated using the 
equation
\begin{equation}
\mathcal{E}_i = \text{Tr} (H_i \rho_i) = \sum_{k} \text{Tr} (H_k^i \rho_k^i),
\end{equation}
with $i=A,~B,~C,~D$.
The quantum Otto cycle works as an engine when energy is absorbed 
from the hot bath ($\mathcal{Q}_{in} >0$), the energy is released to 
the cold bath ($\mathcal{Q}_{out} <0$), and the work is done by the 
engine ($\mathcal{W} <0$), where
\begin{eqnarray}
\mathcal{Q}_{in} &=& \mathcal{E}_B - \mathcal{E}_{A}\\
\mathcal{Q}_{out} &=& \mathcal{E}_D - \mathcal{E}_{C}\\
\mathcal{W} &=& - (\mathcal{Q}_{in} + \mathcal{Q}_{out}).
\end{eqnarray}

We characterize the engine performance using the quantities efficiency and 
power which are computed as
\begin{eqnarray}
\eta = \frac{-\mathcal{W}}{\mathcal{Q}_{in}}\\
\mathcal{P} = \frac{\mathcal{W}}{\tau_{total}}.
\end{eqnarray}

\section{Universal scalings in work output}\label{sec4}

The two unitary strokes of the quantum Otto cycle involve driving  
the Hamiltonian of the WM from one parameter to another. During this driving, 
the quantum critical point may or may not be crossed. Let us quickly revisit
universal scalings in the non-equilibrium dynamics of a quantum 
system which is initially prepared in the ground state of the 
Hamiltonian, and is driven through the critical point linearly with
a speed $1/\tau$. The diverging relaxation time 
close to the CP results in loss of adiabaticity, and thus generation
 of defects (excitations) no matter how slowly the CP is crossed \cite
{PhysRevA.73.063405, RevModPhys.83.863, del2014universality, dutta15quantum}. 
The density of such defects $n_{ex}$ follows a universal 
power law with the rate of driving where the power is determined by the critical 
exponents and dimensionality of the system, and is given by 
\begin{equation}\label{eqn_defectdensity}
n_{ex} \sim \tau^{\frac{-\nu d}{\nu z +1}}.
\end{equation}
Here $n_{ex}$ denotes the defect density, $\nu$ is the exponent associated with correlation length and $z$ is the dynamical exponent with 
$d$ being the dimensionality of the system. This scaling between
the defect density and the rate of driving is called the 
Kibble-Zurek scaling which connects the equilibrium critical 
exponents with the non-equilibrium dynamics. However, this scaling 
gets modified when the driving starts from a thermal equilibrium 
state as opposed to the ground state of the system; 
In Refs. \cite{Polkovnikov2008, PhysRevB.81.224301, PhysRevB.81.012303, PhysRevB.83.094304}, the authors
consider the case when the driving starts from the critical point and obtain a scaling
of defects as a function of temperature and the driving rate, which we present below.

Consider the system at criticality prepared in a thermal equilibrium state corresponding 
to a temperature $T$. This system is then driven far away from the 
critical point 
with a rate $1/\tau$.
Then for fermionic quasiparticles, it has been shown  that the excess number of quasiparticles  excited into the momentum mode $k$ starting from a thermal state at temperature $T$ denoted as $\Delta n_{ex,k}(T)$ is related to the quasiparticles at zero temperature $n_{ex,k}^0$ as \cite{PhysRevB.81.224301, PhysRevB.81.012303, PhysRevB.83.094304}
\begin{equation}
\Delta n_{ex,k}^{T} \sim n_{ex,k}^{0} \tanh(\frac{\epsilon_k}{2T}),
\end{equation}
where  $\epsilon_k$ is the initial energy of the mode $k$. Integrating upto all relevant modes denoted by  $k_{max}$ ($\sim \tau^{-\nu/(z\nu+1)})$ which depends on the rate of driving  (see Ref. \cite{PhysRevB.83.094304} for a detailed calculations), we can calculate the defect density as 
\begin{eqnarray}
\Delta n_{ex}(T) \sim \int_0^{k_{max}} p_k \tanh \frac{\epsilon_k}{2T},
\label{eq_fullnex}
\end{eqnarray}
where $p_k$ is  the two level Landau Zener probability.
While in the limit 
$T \rightarrow 0$, this equation reduces to Eq. \ref{eqn_defectdensity},
the high temperature limit defined by $T \gg \epsilon_k$ can
be obtained by approximating $\tanh(\frac{\epsilon_k}{2T}) \sim \frac{\epsilon_k}{2T}$ and integrating upto $\tau$ dependent maximum $k-$ mode. 
Substituting $\epsilon_k \sim k^z$ for modes near the critical mode followed by integration, the excess defect density starting from a thermal state at temperature $T$ follows
\begin{equation}
\Delta n_{ex} \sim \frac{1}{T} \tau^{\frac{-(d + z)\nu}{\nu z +1}}.
\label{eq_scalingnex}
\end{equation}
One can quantify the non-adiabatic excitations  through the  excess energy  $\mathcal{E}^{excess}$
with respect to the adiabatically evolved state as well.
In case of systems for which 
$\mathcal{E}^{excess}$ is proportional 
to the defect density when far away from the critical point,  such as for the transverse Ising \cite{dziarmaga05dynamics} and X-Y \cite{mukherjee07quenching} models in one dimension and the Kitaev model in two dimensions \cite{sengupta08exact},
similar scalings (Cf. \eqref{eq_fullnex} and \eqref{eq_scalingnex}) hold for excess energy as well. It is to 
be noted the temperature is used only to determine the initial 
thermal state, after which the system follows unitary dynamics during the unitary stroke {\bf D} to {\bf A}.

Now let us move on to discuss how these scalings can be related to the engine parameters. At the end of the non-unitary strokes at 
\textbf{B} and \textbf{D}, the system reaches the thermal states corresponding to 
temperatures $T_H$ and $T_C$, respectively. 
{{We consider $T_H$ to be large so that $B$ is a high entropy state 
which results to $\rho_B \approx \rho_C$ so that $\mathcal{E}_C \approx \mathcal{E}_C^{adia}$ independent of $\tau_1$.}}
On the other hand, the non-adiabatic evolution from \textbf{D} to \textbf{A} due to the 
presence of the critical point and the associated generation of defects increases the 
energy at \textbf{A} which we denote
as \cite{Polkovnikov2008}
\begin{equation}
\mathcal{E}_{A} = \mathcal{E}_{A}^{adia} + \mathcal{E}_{A}^{excess},
\end{equation}
where $\mathcal{E}_A^{excess}$ is the excess energy at \textbf{A}.
Now, the work done is given by 
\ba
\mathcal{W} &=& - (\mathcal{Q}_{in} + \mathcal{Q}_{out})\non\\
&=& -(\mathcal{E}_B - \mathcal{E}_{A} + \mathcal{E}_D - \mathcal{E}_{C})\non\\
&=& -(\mathcal{E}_B - \mathcal{E}_{A}^{adia} - \mathcal{E}_{A}^{excess} + \mathcal{E}_D - \mathcal{E}_{C}^{adia})\non\\
&=& \mathcal{\tilde{W}} + \mathcal{E}_{A}^{excess}
\label{eq:work}
\ea
where $\tilde{\mathcal{W}} = -(\mathcal{E}_B - \mathcal{E}_{A}^{adia} + \mathcal{E}_D - \mathcal{E}_{C}^{adia})$ which is the work output had the evolution from 
\textbf{D} $\rightarrow$ \textbf{A} being fully adiabatic.

Thus the work output upto a constant $\tilde{\mathcal{W}}$ shows 
scaling manifested by $ \mathcal{E}_{A}^{excess}$, i.e.,
\begin{equation}\label{eqn_work}
\mathcal{W} - \tilde{\mathcal{W}} = \mathcal{E}_{A}^{excess}.
\end{equation}


Consider the case where $\alpha_2$ is set to its critical value. 
Extending the scaling results to the Otto cycle in the limit when $T\gg \epsilon_k$, 
we get 
\begin{equation}
\mathcal{E}_{A}^{excess} \sim \frac{1}{T_C} \tau_2^{\frac{-(d + z)\nu}{\nu z +1}}.
\end{equation}
The work output  can be then written as
\begin{equation}\label{eqn_work_gen}
\mathcal{W} - \tilde{\mathcal{W}} = \frac{R_1}{T_C} \tau_2^{\frac{-(d + z)\nu}{\nu z +1}} 
\end{equation}
where $R_1$ is the proportionality constant. From Eq. \ref{eqn_work_gen}, it can be inferred that a wise choice of the WM 
belonging to appropriate universality class and dimensionality greatly helps 
in designing Otto cycles so that it can deliver maximum output work.

As seen from Eq. \eqref{eqn_work_gen}, adiabatic operation of QHE, signified by $\mathcal{W} \to \tilde{\mathcal{W}}$, demands 
\ba 
\tau_2 \gg \tau_{min} =  \left(\frac{R_1}{T_C}\right)^{\frac{\nu z + 1}{\nu(d+z)}}.
\label{eq:tauad}
\ea
Clearly, a higher $T_C$ allows us to achieve adiabatic operation for lower values of $\tau_2$. This can be attributed to the presence of thermal fluctuations at high temperatures, which dominate for $\tau_2 \gg \tau_{min}$. Consequently, increasing $\tau_2$ above $\tau_{min}$ fails to yield any additional work output. 

It is to be noted that Kibble Zurek scalings are valid for $L \gg \hat{\xi} \sim \tau^{\nu/\nu z + 1}$ \cite{dutta15quantum, PhysRevB.86.064304}. Therefore, we expect the expressions given in Eqs. \eqref{eqn_work} - \eqref{eq:tauad} to also hold in this limit, implying the presence of finite size corrections for small system sizes. Notably, Kibble Zurek mechanism in quantum critical systems driven out of equilibrium has been studied experimentally in quantum simulators comprising $256$ atoms \cite{ebadi2021quantum}.

\section{Transverse Ising model as working medium}\label{sec5}
We demonstrate the results derived in the previous section using the prototypical model of transverse Ising model (TIM) as the WM of the Otto cycle. The Hamiltonian of TIM is given by
\begin{equation}\label{H}
H = -J\sum_{n} \sigma_{n}^{z}\sigma_{n+1}^{z} - h \sum_{n} \sigma_{n}^{x}
\end{equation} 
where $J$ is the interaction strength, $\sigma_{n}^{\mu}$ with $\mu = x, y, z$ are the 
Pauli matrices at site $n$, and $h$ is the transverse field which plays the role of 
$\alpha$ in Section \ref{sec2}. The model shows a quantum phase transition from the 
paramagnetic state ($J \ll h$) to the ferromagnetic state ($J \gg h$) at the 
quantum critical point $J = \pm h$ \cite{pfeuty70the, bunder99effect, lieb61two}. We set $J=1$ throughout the paper.

When written in momentum ($k$) space using the basis $|0,0\rangle $, $|k, 0\rangle$, $|0, -k\rangle$, $|k, -k\rangle$, the Hamiltonian takes the form
\begin{equation}
H = \sum_{k>0} \psi_k^{\dagger} H_k \psi_k
\end{equation} 
with $\psi_k^{\dagger} = (c_k^{\dagger}, c_{-k})$ and 
\begin{equation}\label{Ham_new}
H_{k} = \begin{bmatrix}
 -2(h(t) - \cos k) & 0 & 0 & 2 \sin k \\ 0 & 0 & 0 & 0\\ 0 & 0 & 0 & 0\\ 2 \sin k & 0 & 0 & 2(h(t) - \cos k) 
\end{bmatrix}.
\end{equation}
The eigenenergies of $H_k$ are $-\epsilon_k, 0, 0, \epsilon_k$ with $\epsilon_k = 2 \sqrt{(h - \cos k)^2 + \sin k^2}$.

During the unitary strokes of the QOC, the transverse field is changed from $h_1$ to $h_2$ in the \textbf{B} $\rightarrow$ \textbf{C} stroke using the driving protocol
\begin{equation}
h(t) = h_1 + (h_2 - h_1)\frac{t}{\tau_1}, \hspace{0.5cm} 0 < t < \tau_1,
\end{equation}
and vice versa in the \textbf{D} $\rightarrow$ \textbf{A} stroke using the protocol
\begin{equation}
h(t) = h_2 + (h_1 - h_2)\frac{t - a}{\tau_2}, \hspace{0.5cm} a < t < a + \tau_2.
\end{equation}
At the end of the non-unitary strokes, the TIM reaches the thermal equilibrium states corresponding to $h_1$ and $T_H$ at \textbf{B} and $h_2$, 
and $T_C$ at \textbf{D}. The analytical expressions for $\mathcal{E}_B$, 
$\mathcal{E}_C^{adia}$, $\mathcal{E}_D$ and $\mathcal{E}_A^{adia}$ has been calculated 
in the Appendix of Ref. \cite{e24101458} using which the expression for 
$\tilde{\mathcal{W}}$ can be written as
\begin{eqnarray}
&\tilde{\mathcal{W}}&= \sum_k \left( \epsilon_k(h_1)-\epsilon_k(h_2) \right) \nonumber\\
&\lbrace& \frac{\left(e^{-\beta_H \epsilon_k(h_1)} - e^{\beta_H \epsilon_k(h_1)} \right) }{Z(h_1)} \non\\&-& \frac{\left(e^{-\beta_C \epsilon_k(h_2)} - e^{\beta_C \epsilon_k(h_2)} \right)}{Z(h_2)} \rbrace
\end{eqnarray}  
\begin{figure}[h]%
\centering
\includegraphics[width=0.5\textwidth]{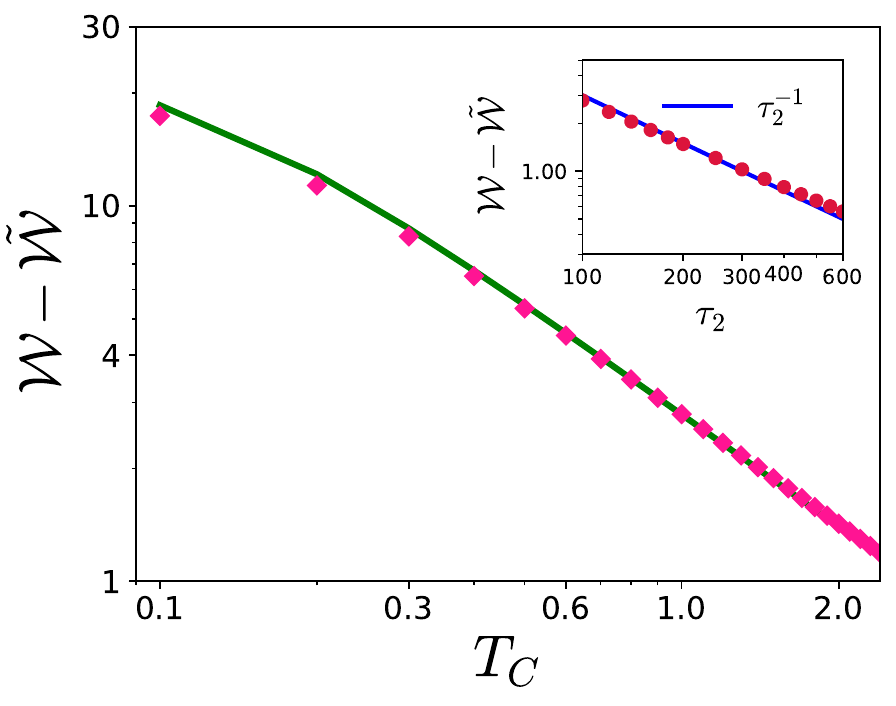}
\caption{The data points correspond to $\mathcal{W}-\tilde {\mathcal W}$ obtained numerically where as the green colored solid line corresponds to the analytical expression obtained by integrating Eq. \ref{eq_fullnex}.
Clearly, one can observe $1/T_C$ scaling at large $T_C$.  The inset shows 
$\mathcal{W} - \tilde{\mathcal{W}}$ as a function of $\tau_2$ for $T_C = 1.0$. The blue solid line corresponds to  $1/\tau_2$.  The parameters used are: $L = 100, h_1 = 10, h_2 =1, T_H = 1000, \tau_1 = 10$.
}
\label{fig_check}
\end{figure}
For TIM, the value of the critical exponents are 
$\nu = z = 1$ which gives
\begin{eqnarray}
\mathcal{W} - \tilde{\mathcal{W}} \sim \frac{1}{T_C \tau_2}.
\end{eqnarray}
One can also obtain the expression for excess defects or excess energy by integrating the analytical expression given in 
Eq. \ref{eq_fullnex} where $p_k$ is given by the 
Landau Zener probability which in our case takes the form
\begin{eqnarray}
p_k= e^{-2\pi \tau_2 \sin^2{k} /(h_1 - h_2)}.
\label{eq_fullnex_tim}
\end{eqnarray}

We plot in Fig. \ref{fig_check} the numerically obtained $\mathcal{W}-\tilde{\mathcal{W}}$, and compare it with the analytical form 
 given in Eqs. \ref{eq_fullnex} and Eq. \ref{eq_fullnex_tim}; as expected, the $1/T_C$
scaling given in Eq. \ref{eq_scalingnex} is satisfied for large $T_c$. 

\begin{figure}[h]%
\centering
\includegraphics[width=0.5\textwidth]{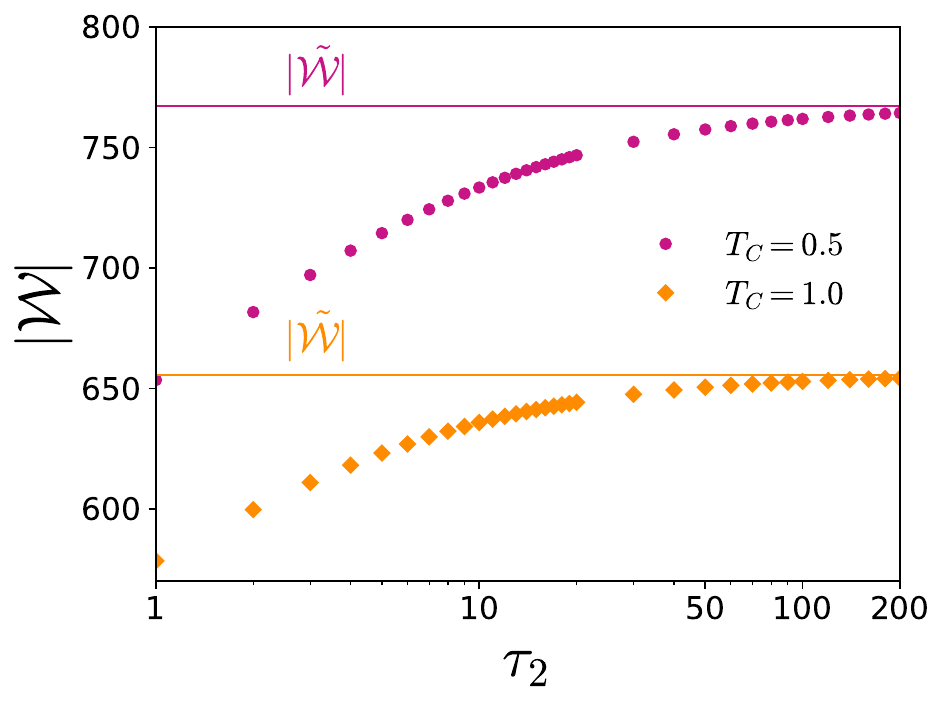}
\caption{$\mathcal{|W|}$ as a function of $\tau_2$ for different values of $T_C$. Here the data points correspond to the numerical values and the solid lines correspond to respective $\mathcal{|\tilde{W}|}$ value. The parameters used are : $L = 100, h_1 = 10, h_2 =1, T_H = 1000, \tau_1 = 10$.}\label{fig_workVstau2_Tc}
\end{figure}
\begin{figure}[h]%
\centering
\includegraphics[width=0.5\textwidth]{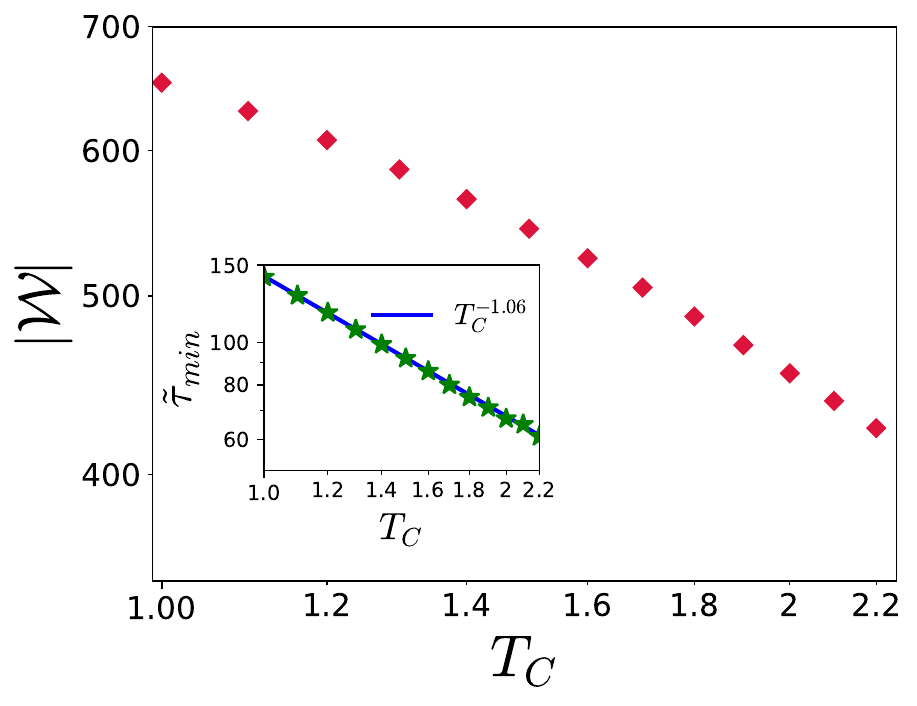}
\caption{$\mathcal{|W|}$ as a function of $T_C$.  The parameters used are : $L = 100, h_1 = 10, h_2 =1, T_H = 1000, \tau_1 = 10, \tau_2 = 100.$ Inset: $\tilde{\tau}_{min}$ as a function of $T_C$ where $\tilde{\tau}_{min}$ is the $\tau_2$ at which $\mathcal{W} - \tilde{\mathcal{W}} < \epsilon$ where  $\epsilon = 2$. The fitted blue continuous line corresponds to a slope of $-1.06$, very close to the theoretical value
of -1 given by Eq. \eqref{eq:tauad}. }\label{fig_workVsTc}
\end{figure}

Let us now focus on the work done $\mathcal{|W|}$. We first show the
presence of $\tau_{min}$, which is the minimum $\tau$ above which 
$\mathcal{W} \to \tilde{\mathcal{W}}$.  Fig. \ref{fig_workVstau2_Tc} gives the plot of $\mathcal{|W|}$ as a function of $\tau_2$ for different values of $T_C$. Clearly, $\mathcal{|W|}$ increases with $\tau_2$ till $\tau_2 \approx \tau_{min}$, after which it saturates. Notably, $\mathcal{|W|}$ saturates at lower $\tau_2$ values for higher $T_C$, as is predicted by Eq. \ref{eq:tauad}. Further,
$|\tilde{\mathcal{W}}|$ is higher for lower $T_C$ as expected, and as also seen in Fig. \ref{fig_workVsTc}.
The inset of Fig. \ref{fig_workVsTc}  shows $\tau_{min}$ as a function of $T_C$, where we have taken $\tau_{min}$ as the $\tau_2$ value for which $\mathcal{W} - \tilde{\mathcal{W}} < \epsilon$, and compared it  with the scaling given by Eq. \eqref{eq:tauad}.

\begin{figure}[h]%
\centering
\includegraphics[width=0.5\textwidth]{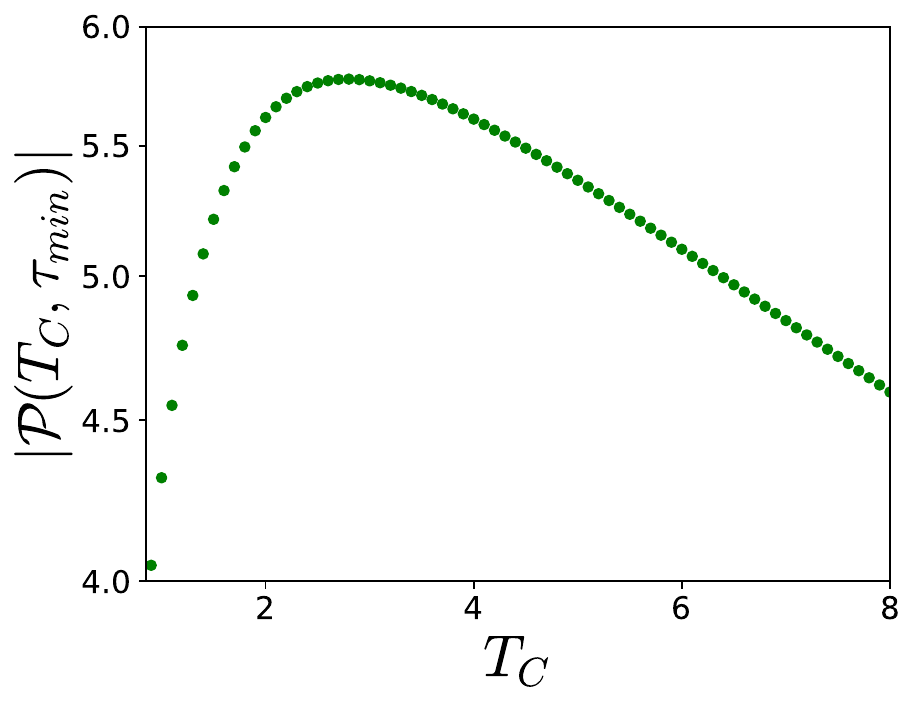}
\caption{$|\mathcal{P}(T_C, \tau_{min})|$ as a function of $T_C$. Here $|\mathcal{P}(T_C, \tau_{min})| = \mathcal{|W|}/ \tilde{\tau}_{min}$ where $\tilde{\tau}_{min}$ and other parameters are same as in Fig. \ref{fig_workVsTc}.}\label{fig_power}
\end{figure}

 In Fig. \ref{fig_power}, we plot the maximum output power $|\mathcal{P}(T_C, \tau_{min})| = |\tilde{W}|/\tau_{min}$ one can obtain without compromising on the work output, i.e., the 
the output power for $\tau_2 = \tau_{min}$, which corresponds to approximately the minimum time for which $\mathcal{W} \to \tilde{\mathcal{W}}$, as a function of $T_C$. Interestingly,  $|\mathcal{P}(T_C, \tau_{min})|$ increases with increasing $T_C$ for small $T_C$, attains a maximum at an intermediate value of $T_C$, before decreasing with increasing $T_C$ for higher $T_C$ values. This can be explained as follows: both $\tilde{\mathcal{W}}$ and $\tau_{min}$ decrease with increasing $T_C$, which eventually results in a peak in the curve. This suggests the intriguing possibility of an optimum cold bath temperature $T_C$ to get high power as well as work output $\mathcal{W} \approx \tilde{\mathcal{W}}$, as  opposed to the zero temperature limit where the work output $\mathcal{W}$ will be maximum ($\mathcal{W} \approx \tilde{\mathcal{W}}$) for $\tau_2 \to \infty$ and $\mathcal{P}(T_C \to 0, \tau_2 \to \infty) \to 0$.

We have assumed that the total time for a single cycle $\tau_{total} \approx \tau_2 = \tau_{min}$. This is especially true for small $T_C$ when $\tau_{min}$ is large (see Eq. 25). Further the rate of evolution during a non-unitary stroke depends on the system-bath coupling strength and the system can be expected to reach infinitesimally close to thermal equilibrium in a finite time.

\section{Conclusion}\label{sec6}
We construct a many body quantum Otto cycle with a WM 
that undergoes a quantum phase transition. The non-unitary 
strokes of the cycle are powered by finite temperature baths, 
while the 
unitary strokes involve driving the WM close to the critical point. 
This driving leads to non-adiabatic excitations which can be 
quantified using relative excess energy that follows universal 
scalings with the rate of driving as well as the temperature of 
the  cold bath. The excess energy can be linked to the output work of 
the engine which thus manifests the universal scalings shown by 
the excess energy. Notably, we show that higher values of the cold bath temperature $T_C$ allows one to operate the engine close to the adiabatic limit for lower values of $\tau_2 \approx \tau_{min}$, which further follows universal scaling relations. This raises interesting questions regarding the importance of control methods such as shortcuts to adiabaticity \cite{hartmann20many}, or bath engineering \cite{e24101458}, for finite temperature quantum heat engines. 
Furthermore, our results for one-dimensional transverse Ising model WM suggest the existence of an optimal value of the cold bath temperature $T_C > 0$,  for operating the QHE with high work output at high power. These counterintituve results stem from the dominance of thermal fluctuations over quantum fluctuations in finite-temperature quantum critical heat engines, for higher bath temperatures.


\backmatter

\bmhead{Acknowledgements} R.B.S. and U.D. acknowledge the use of HPC facility Chandra at IIT Palakkad. U.D. acknowledges support from SERB (SPG/2022/000708). V.M. acknowledges support from SERB through
MATRICS (Project No. MTR/2021/000055) and a Seed
Grant from IISER Berhampur.

\bmhead{Author Contribution Statement} All authors contributed to the study, conception and design. R.B.S. carried out all the analytical and numerical calculations and formal analysis was done by R.B.S., V.M. and U.D. All authors read and approved the final manuscript.
\bmhead{Data Availability Statement} Any data that support the findings of this study are included within the article.

\end{document}